\documentclass[aps,prl,a4paper,twocolumn,showpacs,floatfix,citeautoscript,superscriptaddress,footinbib]{revtex4-1}
\usepackage{graphicx}
\usepackage{amsfonts}
\usepackage{amsmath}
\usepackage{amssymb}
\usepackage{bm}

\usepackage{bbold}
\usepackage{mathtools}
\usepackage{dcolumn}
\usepackage{hyperref}
\usepackage[usenames]{color}
\usepackage{subcaption}

\usepackage{physics}
\usepackage{units}

\usepackage{ragged2e}
\DeclareCaptionJustification{justified}{\justifying}
\newcommand{\out}[1]{{}}

\newcommand{\fref}[1]{Fig.~\ref{#1}}

\begin{document}

\title{Mapping light-dressed Floquet bands by highly nonlinear optical excitations and valley polarization}

\author{Anna Galler}
\address{Max Planck Institute for the Structure and Dynamics of Matter, Center for Free Electron Laser Science, 22761 Hamburg, Germany}
\author{Angel Rubio}
\address{Max Planck Institute for the Structure and Dynamics of Matter, Center for Free Electron Laser Science, 22761 Hamburg, Germany}
\address{Center for Computational Quantum Physics, Flatiron Institute, New York, NY 10010, USA}
\author{Ofer Neufeld}
\address{Max Planck Institute for the Structure and Dynamics of Matter, Center for Free Electron Laser Science, 22761 Hamburg, Germany}

\begin{abstract} 
Ultrafast nonlinear optical phenomena in solids have been attracting major interest as novel methodologies for femtosecond spectroscopy of electron dynamics and control of material properties. Here, we theoretically investigate strong-field nonlinear optical transitions in a prototypical two-dimensional material, hBN, and
show that the $k$-resolved conduction band charge occupation patterns induced by an elliptically-polarized laser can be understood in a multi-photon resonant picture; but remarkably, only if using the Floquet light-dressed states instead of the undressed matter states. Consequently, our work establishes a direct measurable signature for band-dressing in nonlinear optical processes in solids, and opens new paths for ultrafast spectroscopy and valley manipulation.
\end{abstract}

\maketitle

In recent years, strong-field physics and nonlinear optical processes in solids have been intensely investigated~\cite{Ghimire2014,Ghimire2019,sederberg2020attosecond,sederberg2020vectorized,Yue2022,park2022recent}. Such processes allow probing and manipulating ultrafast electron dynamics and material properties with potential attosecond temporal resolution~\cite{langer2016lightwave,baudisch2018,reimann2018subcycle}. For instance, high-harmonic generation (HHG) provides routes for exploring dynamical correlations~\cite{silva2018high,nicolas2018nio,Uchida2022,Murakami2022}, electron-phonon coupling~\cite{Bionta2021,neufeld2022probing}, spectral caustics~\cite{uzan2020attosecond}, exciton formation and dissociation~\cite{freudenstein2022attosecond}, topology~\cite{Bauer2018,silva2019topological,baykusheva2021all,bai2021high,heide2022probing}, and more~\cite{Lakhotia2020,Heinrich2021,Yue2022}. Nonlinear photocurrent generation similarly allows investigating electron coherence and correlations~\cite{higuchi2017light,heide2018,Neufeld2021,heide2022probing}. Specifically in the realm of two-dimensional (2D) hexagonal materials with valley degrees of freedom~\cite{schaibley2016valleytronics,ye2017optical,altarelli2022}, intense femtosecond lasers have been used to non-resonantly control and read the valley pseudospin~\cite{Avetissian2013,Higuchi2017,Langer2018,Galan2020,Mrudul2021,Sharma2022,Silva2022,Dixit2023}, which has technological implications for petahertz electronics, spintronics and memory devices. 

In all of these examples, electronic transitions from the valence to the conduction bands play a pivotal role, e.g. by directly determining the valley polarization, or by forming the essential first step in HHG (both for intra- and inter-band mechanisms) and the photogalvanic effect. Notably, the transition amplitude is determined by the nature of the involved electronic states. Moreover, the specific shape of the bands affects the real-time dynamics of electron propagation through the material. Consequently, it is crucial to ascertain which electronic states are involved - the field-free ones, or the light-dressed ones? The answer to this question is essential not only for fundamental physical understanding, but also for formulating all-optical electronic-structure reconstruction techniques~\cite{vampa2015all,Lanin2017,luu2018measurement,lv2021high} and to approach the idea of Floquet material's engineering~\cite{zhou2023pseudospin,hubener2021engineering,wang2013observation}. Nonetheless, only a few works to date considered possible modifications of the electronic bands in strong-field processes~\cite{Lakhotia2020,Galan2020,Uzan2022,Neufeld2022}, and none in the context of affecting multi-photon transitions from the driving field itself (i.e. without a secondary probe pulse 'sensing' the dressed states). The overwhelming assumption in most works is thus that the field-free bands are a good basis set for interpreting the strong-field dynamics. In this context, previous works also reported a plethora of unique charge excitation patterns upon strong-field driving, which were simply suspected to arise from multi-channel interferences~\cite{Avetissian2013,Higuchi2017,Galan2020,Sharma2022}, but their exact microscopic origin remained unexplained.

Here we theoretically investigate strong-field light-driven excitations in a prototypical 2D material using both sophisticated lattice models and ab-initio time-dependent density functional theory (TDDFT). We explore the direct connection between the electronic structure and the induced conduction band (CB) charge excitation patterns driven by intense elliptically-polarized lasers. We find that the resulting CB occupations do not respect the symmetries of the field-free bands, nor uphold the naively expected energy conservation at multi-photon transitions between field-free states. This unambiguously indicates that laser-dressing plays a major role in the dynamics, even with a single pump field. Interestingly, the CB occupations do follow a clear multi-photon resonant picture if the Floquet light-dressed states are employed instead of the field-free ones. We investigate how this result affects nonlinear valley selectivity. Our work establishes evidence for light-induced electronic structure in nonlinear optics that should be directly detectable with time- and angle-resolved photoelectron spectroscopy (tr-ARPES)~\cite{wang2013observation,soifer2019band,beaulieu2020revealing}, and has significant implications for other nonlinear processes.

\begin{figure}[t!]
\includegraphics[width=\columnwidth]{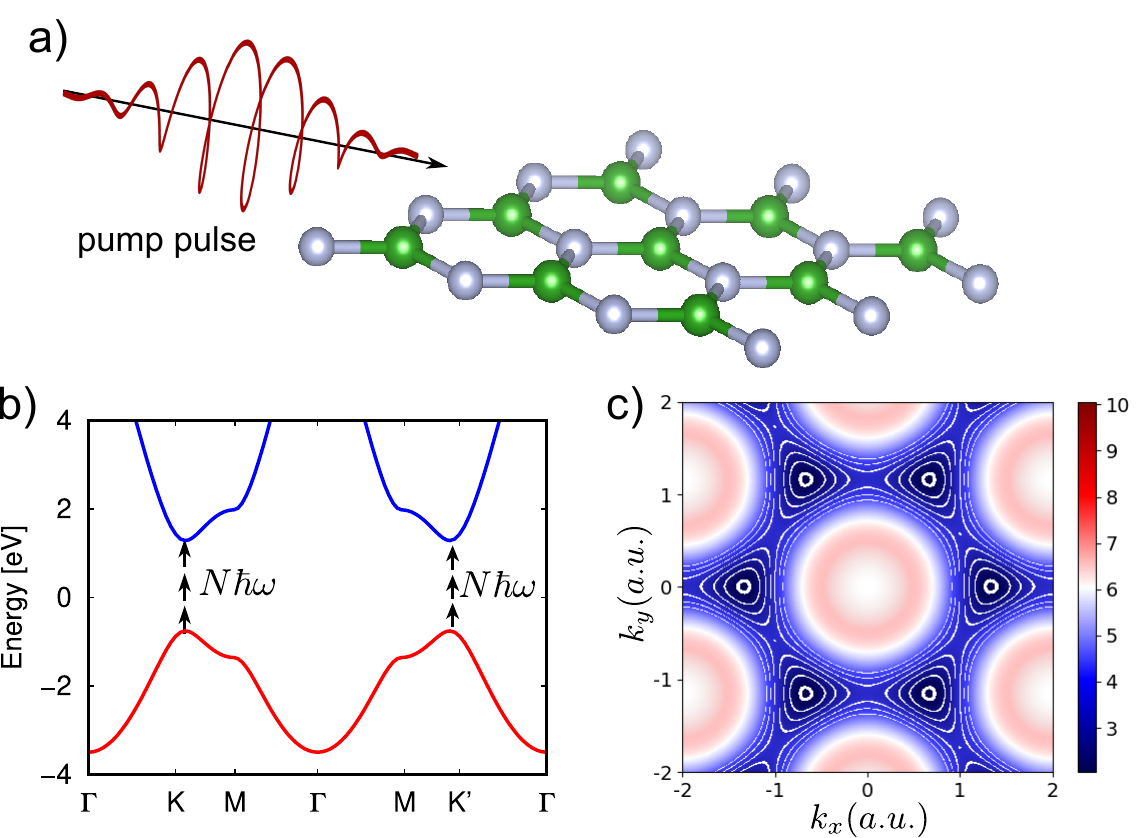}
\caption{(a) Scheme illustration: an elliptically polarized intense pump pulse excites electrons from the valence to the conduction bands. The frequency of the laser pulse is well below the system's band gap, requiring multi-photon transitions (illustrated in (b)). (b) Equilibrium band structure of the employed model along high-symmetry lines (red and blue denoting occupied and unoccupied states, respectively). (c) Equilibrium $k$-resolved band gap of the employed model. The white lines indicate direct gap energies resonant with multiples of the driving photon energy (multi-photon resonant contours).} 
\label{fig:sketch}
\end{figure}

\begin{figure*}[t!]
\begin{minipage}{12. cm}
\includegraphics[width=\columnwidth]{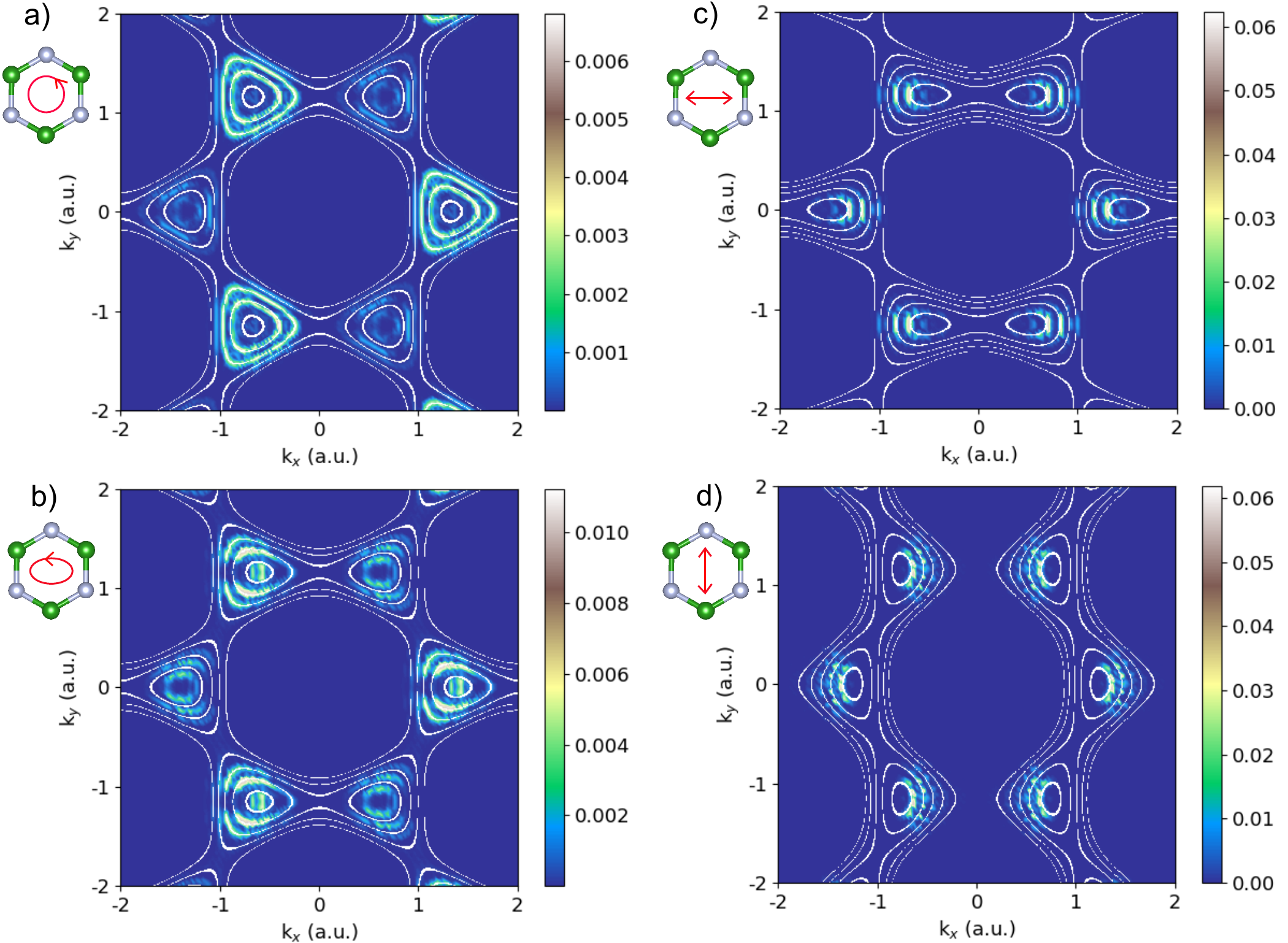}
\caption{Conduction band charge excitation patterns in the model. (a) For a circularly polarized pump pulse, with photon frequency $\unit[0.35]{eV}$ and laser intensity $\unit[0.2]{TW/cm^{2}}$. The white lines indicate energies of the Floquet $k$-resolved direct gap resonant with integer multiples of the driving frequency. (b) Same as (a), but for elliptically polarized driving pulse with $\epsilon=0.6$. (c) Charge excitation pattern for linear driving in $x$-direction. (d) Under linear driving in $y$-direction.} 
\label{fig:model}
\end{minipage}
\end{figure*}

We start by analyzing laser-induced CB charge excitation patterns in a generic two-dimensional model system that exhibits valley degrees of freedom. To this end, we employ a real-space two-dimensional model for a general honeycomb lattice system with dissimilar A/B sublattice sites and periodic boundary conditions, where each site is formed from a local Gaussian potential well (for details on the lattice model see the supplementary information (SI)). The result is a honeycomb lattice with broken inversion symmetry with two electrons per unit cell, leading to a spectrum with a direct optical gap at the $K/K'$ points (see \fref{fig:sketch}). 
Note that each Gaussian locally supports an $s$-like atomic state, but the hybridization leads to nonzero electronic angular momenta and Berry curvature in both the $K/K'$ valleys. Thus, this is the simplest possible real-space model for a system with valley degrees of freedom, where only one dominant valence and conduction bands play a role. 
For our chosen parameters (an $8\%$ difference in the potentials of the A/B sites), we obtain a direct optical gap of $\sim$$\unit[2]{eV}$ at $K$ and $K'$ (see \fref{fig:sketch}b). We simulate the interaction of this electronic system with an intense elliptically polarized laser pulse (up to powers of $\sim$0.2 TW cm$^{-2}$), with a non-resonant carrier frequency that is well below the band gap. The numerical methodology consists of solving the time-dependent Schr\"odinger equation for the electronic dynamics (assuming the dipole approximation for the light-matter coupling term), while assuming the independent particle approximation (i.e. neglecting electron-electron interactions). All technical details of the simulations are delegated to the SI. Importantly, due to the non-resonant conditions, highly nonlinear optical processes involving multiple photons are required to excite electrons from the valence to the conduction bands (see illustration in \fref{fig:sketch}b). The resulting charge distribution patterns in the CB after the laser pulse ends are calculated by projecting the final states onto the field-free states. ~\fref{fig:model} presents such exemplary spectra for several driving conditions, showing the emergence of distinct ring-like patterns. 

Let us first analyze the induced occupations under circular driving (\fref{fig:model}a). Here the ring-shaped patterns form a triangular shape reminiscent of the trigonal wrapping of the Brillouin zone (BZ) in the field-free bands of the honeycomb system, e.g. as reflected in the $k$-dependent gap in \fref{fig:sketch}c. Thus, our initial suspicion is that the shape of the ground-state bands is imprinted onto the $k$-space occupation patterns induced by the external light field. Before exploring this hypothesis, it is important to note that, within the dipole approximation and our employed methodology, only direct optical transitions that conserve the electron crystal momentum are considered. 
Thus, one expects that the transition amplitude and final CB occupations $g_{CB}$, would be determined by a Fermi's golden rule expression of the type $g_{CB}(\textbf{k})\sim|\int_{0}^{t_f} dt \bra{\psi_{\textbf{k},v}}V_{int}(t)\ket{\psi_{\textbf{k},c}}|^2$, where $V_{int}$ is the laser-matter interaction, $\ket{\psi_{\textbf{k},v/c}}$ represents the field-free Bloch-state at $\textbf{k}$, and $t_f$ is the time at which the laser pulse ends. Importantly, since the laser is monochromatic, the integral 'selects' energies that are resonant with a multi-photon transition condition, i.e. at $k$-points that uphold $\epsilon_{c}(\textbf{k})-\epsilon_{v}(\textbf{k})=n\omega$, where $n$ is any integer (equivalent to an energy conservation condition). However, if we overlay the resonant multi-photon lines obtained from the ground-state system (\fref{fig:sketch}c) onto the resulting occupation patterns (\fref{fig:model}a), they do not match---only the expected qualitative shape of the triangular wrapping is reproduced, while energies are shifted in some conditions by a maximal displacement of $\sim0.5\omega$ (see SI). The situation worsens if we consider other driving polarizations where the trigonal wrapping is lost (\fref{fig:model}b-d). Overall, it is obvious that this multi-photon picture is too simplistic. It is also clear that the symmetries of the laser-matter system play a role in determining the final electronic occupation patterns. 

At this stage, we consider an alternative explanation for the CB occupation patterns that relies on a laser-dressing picture for the electronic system. From a formal perspective, the coupling of the electronic system to the laser in the model is performed with a Peierls substitution~\cite{graf1995electromagnetic,moos2020intense}. Thus, one expects the electron momenta to couple to the laser vector potential $\textbf{A}(t)$, and $\textbf{k}\to\textbf{k}(t)=\textbf{k}-\textbf{A}(t)/c$. Consequently, the field-free band structure can be thought of as effectively 'shaking' in time along the laser polarization axis, opening up other resonant multi-photon channels between different regions in $k$-space. The resulting transitions should reflect a complicated time-average of those available channels, weighted by the particular intensity of the laser and density of states in each moment in time. Such a picture would allow rotational symmetry breaking and should shift the peaks from the ground-state resonant conditions, as observed numerically. Mathematically however, it is not clear how the averaging procedure should be performed.

\begin{figure*}[t!]
\begin{minipage}{17. cm}
\includegraphics[width=\columnwidth]{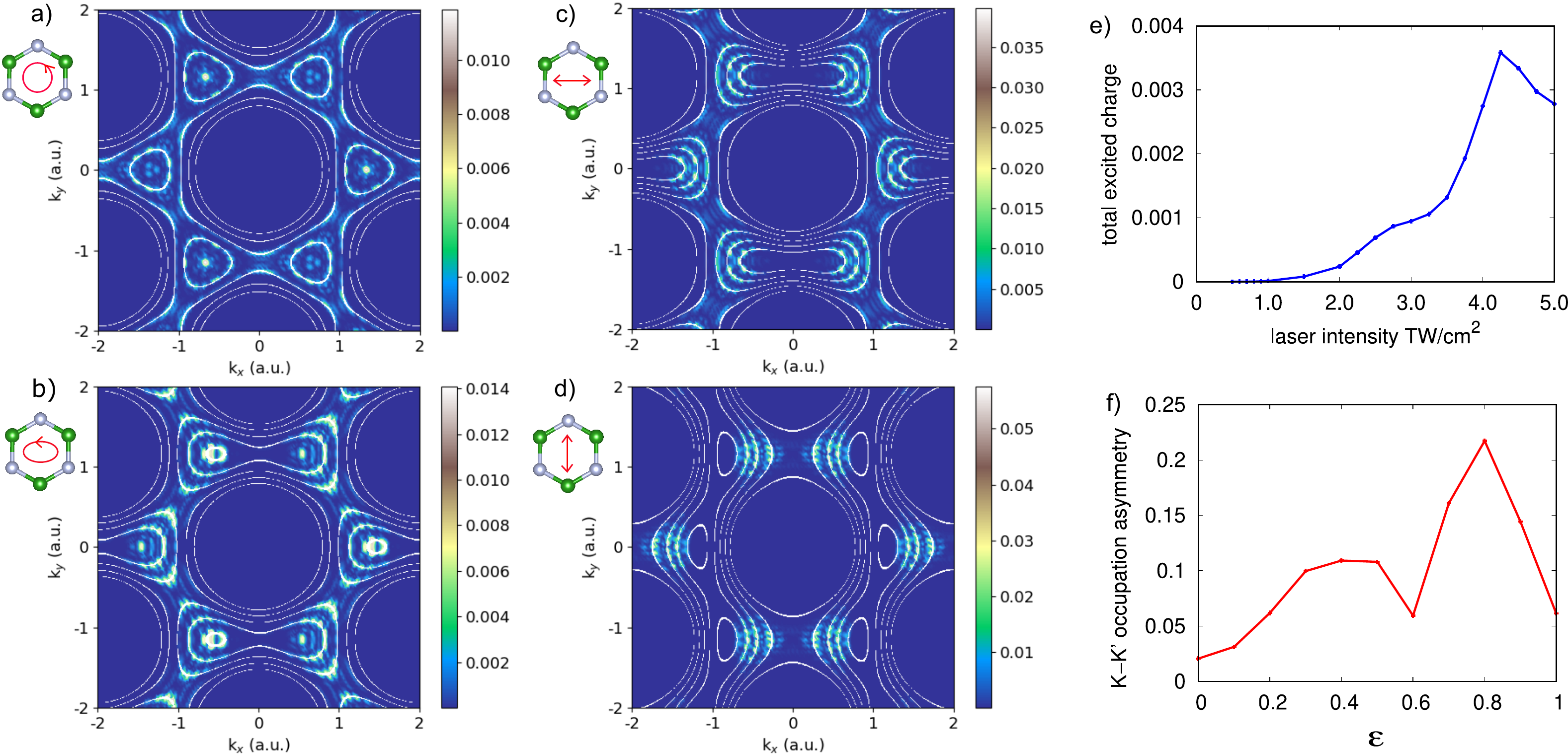}
\caption{Charge excitation patterns and highly nonlinear valley asymmetry in hBN. (a) Conduction band charge excitation pattern under circular driving, (b) an elliptically polarized pump pulse with $\epsilon=0.6$, (c) a linearly polarized pulse in $x$-direction and (d) linear driving in $y$-direction. In all depicted cases, the laser intensity is $\unit[2]{TW/cm^{2}}$ and the photon frequency $\unit[0.7]{eV}$. The white lines indicate energies of the Floquet $k$-resolved direct gap resonant with integer multiples of the driving frequency. (e) The total excited charge in the conduction bands vs. laser intensity. (b) Valley polarization in hBN vs. driving ellipticity (for laser power of $\unit[2]{TW/cm^{2}}$).} 
\label{fig:hBN}
\end{minipage}
\end{figure*}

An alternative but equivalent description can be achieved with Floquet theory, where the 'shaken' bands are replaced by the quasi-energy bands~\cite{Holthaus2016}. The quasi-energy bands are eigenstates of the quantum propagator and have constant occupations, allowing non-ambiguous description of the dynamics. This is the standard approach for describing laser-dressing effects in many time-periodic systems (see e.g. refs.~\cite{oka2019floquet,bao2022light,rudner2020band}). Thus, a natural extension of the hypothesis above is to simply replace the ground-state Bloch states in Fermi's golden rule with Floquet-Bloch states, generating a resonant condition at $k$-points that uphold $\epsilon^{F}_{c}(\textbf{k})-\epsilon^{F}_{v}(\textbf{k})=n\omega$, where $\epsilon^{F}_{c/v}(\textbf{k})$ are the Floquet quasi-energy bands, which are light-dressed and differ from $\epsilon_{c/v}(\textbf{k})$. Let us emphasize that it is \textit{a priori} not clear whether or not such a replacement is legitimate. Even if the Floquet states are eigenstates of the driven system, it is not obvious that one could formulate a Fermi's golden rule with them, because: (i) the states are time-dependent, (ii) the interaction with the laser is already incorporated into the states themselves, (iii) there are ambiguities in determining the ordering and precise eigenenergies of the states, and (iv) Fermi's golden rule arises from time-dependent perturbation theory, whereas the Floquet states are non-perturbative entities.
Nevertheless, to further test this hypothesis, we construct a two-band tight-binding (TB) Hamiltonian $H(\textbf{k})$ for the model, including up to 5'th order nearest-neighbour (NN) hopping terms (following ref.~\cite{wang2013observation}). 
All hopping terms were fitted such that $H(\textbf{k})$ reproduces the bands obtained from the real-space model (see SI for details). The TB Hamiltonian is expected to correctly capture the generic electron dynamics induced by the laser field, while deviations are only expected due to the Hamiltonian not including the full real-space dynamics of Bloch states. We couple $H(\textbf{k})$ to an external laser through a Peierls substitution and calculate the corresponding Floquet quasi-energy bands of the light-driven Hamiltonian assuming perfect temporal periodicity (neglecting the laser envelope~\cite{Neufeld2019}). The resulting time-dependent Hamiltonian $H(\textbf{k}(t))$ is decomposed into harmonics of $\omega$ to obtain the sub-blocks of the Floquet Hamiltonian:
\begin{equation}
\label{eq:floquet}
H_{F}^{n,m}=\delta_{n,m}n\omega\sigma_0+\frac{\omega}{2\pi}\int_{0}^{\frac{2\pi}{\omega}} H(\textbf{k}(t))e^{i\abs{n-m}\omega t} \,dt 
\end{equation}

where $|n-m|$ is the photon channel order and the integrals are numerically solved for each $k$-point. Subsequently, the Floquet Hamiltonian is diagonalized to obtain the quasi-energies, which are corrected by their photon-channel index. The resulting light-dressed bands, $\epsilon^{F}_{c/v}(\textbf{k})$, are taken as the bands that converge to the correct field-free bands in the limit of zero laser power.

\fref{fig:model} presents the main result of this Letter---it compares the numerically obtained charge excitation patterns (i.e. from directly solving the time-dependent Schr\"odinger equation for the real-space model coupled to laser driving) with multi-photon-resonant energy contours in the light-dressed bands (overlaid in white). The two match very well in all of the examined laser regimes, including different laser wavelengths, powers, and polarizations (see SI). Thus, the numerical results validate the hypothesis above. We further discuss some noteworthy points: (i) The exceptionally good matching between the CB occupations and the resonant multi-photon transition picture in the light-dressed system means that the driving laser both dresses the system, and induces the optical transitions. Consequently, our results propose an observable that is directly sensitive to light dressing without an additional probe pulse. (ii) In the case of a circularly-polarized laser driving (\fref{fig:model}a), the triangular-shaped rings around $K/K'$ reflect the shape of the Floquet bands---since the circular pulse respects the 3-fold rotational symmetry of the lattice~\cite{alon1998selection,neufeld2019floquet}, trigonal wrapping is preserved. (iii) The $K/K'$ valleys couple differently to circularly-polarized components of the laser due to the valley degrees of freedom, and the direct gaps at $K$ and $K'$ differ. Crucially, the latter is the origin of the different charge excitation patterns in the $K$ and $K'$ valleys in this non-resonant intense driving regime. Together with optical selection rules based on the orbital angular momentum in the valleys~\cite{sengupta2018optically,cheng2019chiral}, this effect gives rise to valley asymmetry in the highly nonlinear non-resonant regime. It's also noteworthy that in the high-frequency and/or weak-driving limits, the Floquet bands coincide with the field-free bands, and the standard Fermi golden rule picture is restored (the Floquet picture can be considered a generalization of the standard approach). (iv) For generic elliptical driving the CB occupations become compressed along or transversely to the driving axis. This provides an all-optical knob for tuning the valley selectivity.

Next, we validate this model result in a realistic 2D material. We perform ab-initio calculations for a monolayer of hexagonal boron-nitride (hBN) irradiated by an intense laser pulse with a frequency of $\unit[0.7]{eV}$, well below the band gap of hBN (\unit[4.2]{eV} with the local density approximation). We employ a real-time TDDFT approach as implemented in the Octopus~\cite{Octopus2006,Octopus2015,Octopus2020,hartwigsen1998relativistic} code. The methodology is similar to the one employed for the model above, but with multiple optically-active valence electrons that interact with each other as well as with the driving laser (for details see SI). ~\fref{fig:hBN}a-d shows the corresponding CB excitation patterns for several driving conditions. The patterns are overlaid with the multi-photon-resonant transition contours obtained for the Floquet quasi-energy bands from a TB model 
with the TB parameters fitted to the hBN first valence and conduction bands. Overall, the induced patterns agree remarkably well with the Floquet resonant transitions and effectively map the light-dressed bands. Small deviations can be observed here because the TDDFT calculations include excitations from multiple valence bands to multiple conduction bands, as well as electron-electron interactions, both of which are neglected in the Floquet TB approach.

~\fref{fig:hBN}f further shows a quantitative measure for the valley asymmetry in hBN, defined as $P=(n_K-n_{K'})/(n_K+n_{K'})$, where $n_K$ ($n_{K'}$) are the electron occupations in each valley obtained by integrating the population around $K$ ($K'$) (see SI). 
The valley polarization increases with the driving ellipticity $\epsilon$. However, the increase is not monotonic, unlike in the resonant case~\cite{sengupta2018optically}. We believe that this is testament of the complex CB occupation patterns obtained in the highly nonlinear regime, also reflecting the complex structure of the Floquet bands. Indeed, \fref{fig:hBN}e shows that the total CB excitation in these conditions is a non-perturbative non-linear observable. Thus, our results could form a new approach for analyzing and predicting valley asymmetry under strong-field and non-resonant driving.

To summarize, we have investigated strong-field nonlinear optical excitations in 2D materials. We found that the excitations can be analyzed within a multi-photon-resonant picture, but only if the Floquet light-dressed states are employed for calculating the transition energies. Importantly, the $k$-space distribution of electrons in the conduction and valence band after the laser pulse effectively map the light-driven Floquet bands, including potential symmetry breaking induced by the laser. These phenomena are directly experimentally accessible by tr-ARPES~\cite{Neufeld2021,neufeld2022time}.
While we only explored here 2D hexagonal materials, the results should be general to all periodic systems. Looking forward, since our findings establish a clear connection between the light-dressed electronic structure and the material's nonlinear optical excitations, they should affect research in all connected fields such as HHG, harmonic side-band generation, and nonlinear photogalvanic effects, implying possible new interpretations of ultrafast spectroscopies based on these techniques. 

{\em Acknowledgments.} 
We thank Massimo Altarelli, Andrey Geondzhian, Wenwen Mao and Shunsuke Sato for helpful discussions. This work was supported by the Cluster of Excellence Advanced Imaging of Matter (AIM) – EXC 2056 - project ID 390715994, SFB-925 "Light induced dynamics and control of correlated quantum systems", project 170620586 of the Deutsche Forschungsgemeinschaft (DFG), Grupos Consolidados (IT1453-22), and the Max Planck-New York City Center for Non-Equilibrium Quantum Phenomena. The Flatiron Institute is a division of the Simons Foundation. O.N. gratefully acknowledges the generous support of a Schmidt Science Fellowship.

%

\end{document}